\documentclass[twocolumn,superscriptaddress,floatfix,preprintnumbers,prl ,nofootinbib,hyperref]{revtex4-2}
\usepackage[colorlinks=true,breaklinks=true]{hyperref}
\usepackage[utf8]{inputenc}
\hypersetup{allcolors=[rgb]{0.0 0.0 0.6},linkcolor=[rgb]{0.75 0.05 0.05}}
\usepackage{mathtools}
\usepackage[dvipsnames]{xcolor}
\usepackage{float}
\usepackage{amsmath}
\usepackage{empheq}
\usepackage[theorems,skins]{tcolorbox}
\usepackage{letltxmacro}
\LetLtxMacro{\oldcite}{\cite}
\renewcommand{\cite}[1]{\mbox{\oldcite{#1}}}
\newtcolorbox{mymathbox}[1][]{colback=white, sharp corners, #1}

\usepackage{orcidlink}

\long\def\exclude#1{}

\newcommand{\beq}{\begin{equation}}
\newcommand{\eeq}{\end{equation}}

\def\ga{\,\,\raise0.14em\hbox{$>$}\kern-0.76em\lower0.28em\hbox
{$\sim$}\,\,}

\allowdisplaybreaks

\bibpunct{[}{]}{,}{n}{}{,}

\begin{document}

\hfill CERN-TH-2023-092

\title{A new source for light dark matter isocurvature  in low scale inflation}
%\title{Axion dark matter can challenge quantum gravity}

\author{Andrea Caputo} %\email{andrea.caputo@uv.es}
\affiliation{Theoretical Physics Department, CERN, 1211 Geneva 23, Switzerland}

\author{Michael Geller}
\affiliation{School of Physics and Astronomy, Tel-Aviv University, Tel-Aviv 69978, Israel}

\author{Giuseppe Rossi} %\email{raffelt@mpp.mpg.de}
\affiliation{School of Physics and Astronomy, Tel-Aviv University, Tel-Aviv 69978, Israel}

%\date{January 25, 2022, %{\color{red}For Resubmittal \today}}

%==========================

\begin{abstract}

Light scalar and pseudoscalar particles are compelling dark matter candidates, with a vast running experimental program to discover them. Previous studies have shown that these light fields can generate sizable isocurvature perturbations in high scale inflationary models. Thereby, dark matter existence and cosmic microwave background measurements impose an upper bound on the inflationary scale.
%In recent years, however, appealing to quantum gravity arguments, it was conjectured that the inflationary scale may actually be extremely low. 
In this work, focusing on the axion case, we point out that light fields present during inflation can generate important isocurvature perturbations also in scenarios of low-scale inflation. In our mechanism, the axion field starts with some non-zero field value during inflation and rolls along its potential. Since inflation has a different duration in different patches of the universe, different regions will then have different values of the axion field, generating cold dark matter isocurvature modes. These modes are fully correlated with the adiabatic ones and share the same spectral index. In this scenario, the axion mass determines a lower bound on the Hubble parameter during inflation. 

\end{abstract}

\maketitle

%%%%%%%%%%%%%%
%%%%%%%%%%%%%%

\section{Introduction} 
The particle nature of dark matter and the physics of cosmic inflation are two of the biggest open questions in cosmology. For one of the standard dark matter candidates, the axion\cite{Peccei:1977hh, Weinberg:1977ma, Wilczek:1977pj, Preskill:1982cy}, these two topics are intricately related. 
%\giuseppe{I would change these two lines into: The particle nature of dark matter and the origin of cosmological structure are two of the biggest open questions in cosmology. Cosmological observations give us the unique opportunity to test different dark matter models, placing non trivial constraints on their parameter space. On the other hand, the existence of a certain dark matter candidate can restrict the parameter space for comsological models. 
%In this work, we investigate the possible signature of light dark matter fluctuations in the context of low scale inflation. 
The axion is known to leave its imprints in primordial fluctuations that seed Cosmic Microwave Background (CMB) anisotropies and maps Large Scale Structure (LSS). If the axion is produced by  the misalignment mechanism, its relic abundance today is tied to the value of the axion field at the end of inflation, which is misaligned from its minimum. 
Therefore, in typical scenarios of high-scale inflation, the axion dark matter abundance is sensitive to de Sitter quantum fluctuations which add a perturbation to the initial value of the axion field~\cite{Beltran:2006sq, Graham:2018jyp, Linde:1991km, Feix:2019lpo}. These fluctuations do not observably perturb the 
total energy density during inflation, but instead give rise to dark matter \textit{isocurvature} perturbations in the absence of thermal symmetry restoration after inflation. The CMB is sensitive to the presence of such fluctuations, which cannot be reduced to the fluctuations of single clock. Planck data therefore set a limit on the axion decay constant, $f_a$ and -- in the case of axion discovery -- would determine an \textit{upper} bound on the inflation scale.

In this work we show that the production of isocurvature in axion models is not limited to this scenario of high scale inflation. In particular, we identify a new source of isocurvature perturbation in axion dark matter models for low scale inflation. While usually inflationary scale is taken to be very large, the possibility of low scale inflation is also plausible and allowed by data. Interestingly, low-scale inflation scenarios may even be \textit{preferred} by quantum gravity theories~\cite{Bedroya:2019tba}. Previously, isocurvature bounds were only considered as an \textit{upper} bound on the scale of inflation. Here, we show that a \textit{lower} bound on the scale of inflation exists as well. In face, sizable isocurvature perturbations arise if the axion field traverses some distance along its potential during inflation. This occurs when the Hubble scale during inflation is close to the axion mass, implying that the lower bound on the inflation scale is around the weak scale. While we focus for simplicity and concreteness on axions, the mechanism we study is relevant for any light scalar or pseudoscalar dark matter candidate produced via the misalignment mechanism. 

The paper is organized as follows. We start with an intuitive description of the mechanism and the expected scaling of the effect with the axion mass. We then introduce low-scale inflation and some of its features and theoretical motivations. Next, we compute the adiabatic and isocurvature perturbations in a simple model of two fields, the inflaton and the axion. We then derive constraints on the inflationary scale from CMB measurements and compute the expected adiabatic and isocurvature spectral indices. Finally, we summarize and conclude.

\section{Mechanism}
We can intuitively understand the mechanism using a separate universe approach. On super-horizon scales, different patches of the universe will behave like separate FRW universes with different parameters. Because of quantum fluctuations, inflation will last a different amount of e-foldings in different patches, producing locally different scale factors. The curvature (or adiabatic) perturbation is related to the fluctuation in the number of e-foldings via $\delta N = \mathcal{R}$. 
In our scenario, the axion field starts with some non-zero field value during inflation and traverses some non-negligible distance towards the minimum of its potential. This distance depends on the duration of inflation.  
Since inflation has a different duration in different patches, regions that have more (less) time to inflate will have a bigger (smaller) scale factor, therefore the value of the axion field at the end of inflation will be locally smaller (larger), producing axion fluctuations that are correlated with the curvature fluctuations. These field fluctuations will produce perturbations in the dark matter energy density once the axion field starts oscillating after inflation.
The perturbations produced in this way are not adiabatic, i.e. different patches will not only have a different scale factor (and hence a different energy density) but also different dark matter concentration. These perturbations are called isocurvature and they are constrained by CMB data at large angular scales. Using this picture it is also easy to estimate the size of the effect. At the end of inflation each patch will have 
\begin{align}
\delta \phi_a = \frac{d \phi_a}{dN} \delta N = \frac{\dot{\phi}_a}{H} \mathcal{R} 
\end{align}
% The time translation that removes the inflaton perturbation is $\delta N \sim \frac{H\delta \phi} {\dot{\phi}} \sim \frac{H^2}{\dot{\phi}}\sim  \Delta_R  $. This translation induces an axion perturbation $\delta a\sim \frac{\dot{a}}{H}\delta N$, we find that 
and, as a consequence 
\begin{equation}
\delta_{\phi_a}\sim 2\frac{\delta \phi_a}{\phi_a}\sim 2 \frac{\dot{\phi_a}}{H \phi_a}\Delta_R\sim \frac{2 \, m_a^2}{H^2} \mathcal{R}. 
\end{equation} 

Summing up, the core of our effect is the following: we have two fields during inflation, \textit{both of them rolling down} their potentials. For this reason, two clocks are present during inflation and isocurvature fluctuations are expected. In fact, the inflaton perturbations alone only change the inflaton clock and therefore create a mismatch between the two clocks, thereby generating axion isocurvature. Therefore, on one hand our effect is relevant only when the axion field is massive enough because it needs to roll. On the other hand, the axion cannot be too massive, otherwise the relic abundance (and consequently the size of the perturbations) will be exponentially suppressed. 

\section{Inflationary Dynamics} 

\subsection{Low Scale Inflation}

We start by providing a brief overview of low-scale inflation and typical parameters of interest. Low-scale inflation has gained recent interest in the context of quantum gravity, in particular by the Trans-Planckian Censorship Conjecture (TCC)~\cite{Bedroya:2019tba}, which limits the number of e-foldings during inflation and predicts the Hubble parameter during inflation to be below $\rm GeV$, or much smaller in the case of single field inflation $H \lesssim 0.01$ eV~\cite{Kadota:2019dol}. A tighter upper limit is also obtained if~\cite{Brandenberger:2019eni} the universe
is radiation-dominated (and expanding) between the
Planck time and the onset of inflation.
One peculiarity of these models is an extremely small value of the first inflationary parameter
\begin{equation}
\epsilon = \frac{H^2}{8 \pi^2 M_{\rm pl}^2 \Delta_{\mathcal{R}}^2} \approx 10^{-49} \Big(\frac{H}{\rm eV}\Big)^2,
\end{equation}
where we have used the CMB normalization $\Delta_{\mathcal{R}}^2 \simeq 2.2 \times 10^{-9}$~\cite{Planck:2018vyg}. An extremely small $\epsilon$ also leads to an extremely small tensor to scalar ratio, $r \simeq 16 \, \epsilon$. \footnote{Notice that in multifield models, the tensor to scalar ratio is fixed at horizon crossing and not at the end of inflation \cite{Wands:2002bn}. This means that in our scenario it would be dominated by the axion slow roll parameter rather than the inflaton.}  
This also means that any detection of primordial gravitational waves on cosmological scales would rule out these models.

Another peculiarity is a very small reheating temperature
\begin{equation}
T_{\rm reh} \equiv \Big(\frac{g_*\pi^2}{90}\Big)^{-1/4}\sqrt{H \, M_{\rm Pl}} \simeq 10^5 \, \rm MeV \sqrt{H/10^{-6} \rm eV},
\end{equation}
which can however be safely larger than actual direct bounds (which set $T_{\rm reh} \gtrsim 4.7 \, \rm MeV$~\cite{deSalas:2015glj}). While in many baryogenesis models, 
the reheating temperature must be above the electroweak scale to produce the baryon asymmetry, we note that baryogenesis can still be obtained for lower values of $T_{\rm reh}$~\cite{Pierce:2019ozl, Grojean:2018fus, Dimopoulos:1987rk, Elor:2018twp}.

Notice that in low scale inflation the required number of e-folds to solve the horizon problem is: 
\begin{align}
N > \log\left(\frac{T_0}{T_R} \right) \sim 30.
\end{align}

For simplicity, we assume that the only fields lighter than the Hubble scale are the axion ($\phi_a$) and the inflaton ($\phi_i$) for which we take a simple slow roll potential. Our theory can be described by the action:
\begin{align}
S = \int d^4 x \sqrt{-g} \left(\frac{M_{Pl}^2}{2} R + \sum_{j=i,a} \frac{1}{2} \partial_\mu \phi_{j} \partial^{\mu} \phi_j - V_j\right). 
\end{align}
These fields are driving the system to a phase of quasi-de Sitter exponential expansion with slow-roll parameters:
\begin{align}
    \epsilon_j =& \frac{M_{Pl}^2}{2} \left( \frac{V'_j}{V}\right)^2 \\ 
    \eta_j & =  M_{Pl}^2\frac{V''_{j}}{V}.
\end{align}
We will further assume that the energy density is dominated by the inflaton field $V_i \gg V_a$.

\subsection{Inflationary Perturbations}
One of the most important aspects of inflation is that it provides a  quantum mechanical mechanism for the generation of the fluctuations observed in the cosmic microwave background and in the large scale structure. These fluctuations are believed to be the vacuum fluctuations of the light fields active during inflation, as we now review. 
The perturbed metric, involving only scalar degrees of freedom in the Newtonian or longitudinal gauge reads \cite{Mukhanov:1990me, Polarski:1994rz}:
\begin{align}
ds^2 = (1+2 \Phi) dt^2 - a(t)^2 (1-2 \Psi) d\vec{x} \cdot d\vec{x}
\end{align}
And in absence of anisotropic stress $\Phi = \Psi$ \cite{Mukhanov:1990me}. 
The perturbations follow Einstein's equation
\begin{align}
\dot{\Phi}+H \Phi=4 \pi G \left(\sum_i \dot{\phi}_i \delta \phi_i \right) 
\end{align}
 and the Klein Gordon equation for each field:
 \begin{align}\label{eq:perturbations}
\delta \ddot{\phi}_{j}+3 H \delta \dot{\phi}_{j}+\left(\frac{k^{2}}{a^{2}}+V_{j}^{\prime \prime}\right) \delta \phi_{j}=4 \dot{\phi}_{j} \dot{\Phi}-2 V_{j}^{\prime} \Phi,
\end{align}
together with the constraint equation:
\begin{align}
\left(\dot{H} + \frac{k^2}{a^2} \right) \Phi = 4 \pi G_N \sum_j \left(-\dot{\phi}_j \delta \dot{\phi}_j + \ddot{\phi}_j \delta \phi_j \right). 
\end{align}

During inflation the scale factor exponentially increases while the Hubble parameter $H$ is decreasing, so at very early times all modes of interest are subhorizon $q \equiv k/a \gg H$. According to the equivalence principle, the dynamics of the scalar fields $\delta \phi_j$ at short scales must match that of a massless scalar field in Minkowski space-time. 
The typical assumption is that the field fluctuation are in their vacuum state of this Minkowski space-time. This is the so called called Bunch-Davies state. They take the form  
\begin{align}
\delta \phi_j(k, t) = \frac{H(t_k)}{\sqrt{2 k^3}} e_j (\mathbf{k})
\end{align}
where $k = a H $ and $e_i(\vec{k})$ are Gaussian random variables satisfying $\left\langle e_{j}(\mathbf{k}) e_{j^{\prime}}^{*}\left(\mathbf{k}^{\prime}\right)\right\rangle=\delta_{j j^{\prime}} \delta^{(3)}\left(\mathbf{k}-\mathbf{k}^{\prime}\right)$~\cite{Polarski:1994rz}. 
The field fluctuations will maintain approximately frozen until the modes become super-horizon and the equations of motion can be approximated as: 
\begin{align}
    3 H \delta \dot{\phi_j} + V''_j \delta \phi_j = -2 V'_j \Phi. 
\end{align}
These equations can be integrated analytically, see Ref.~\cite{Polarski:1994rz} for the original derivation and the Supplementary material. 
The non decreasing modes are:
\begin{align}\label{EqMaster1}
    \Phi  =-C_{1} \frac{\dot{H}}{H^{2}} +& \frac{C_3}{3} \frac{V_i V^{\prime 2}_a + V_a V^{\prime 2}}{(V_a+V_i)^2} \nonumber\\
\frac{\delta \phi_{i}}{\dot{\phi}_{i}} =\frac{C_{1}}{H} +& 2 \,H  C_3 \frac{ V_a}{V_a+V_i} \\
\frac{\delta \phi_{a}}{\dot{\phi}_{a}} =\frac{C_{1}}{H} -& 2 \, H C_3 \frac{ V_i}{V_a + V_i} \nonumber.
\end{align}

Matching these equations to the fields at horizon crossing gives:
\begin{align}
&C_{1}(\mathbf{k}) = -\frac{8 \pi G H_k}{\sqrt{2 k^{3}}} \sum_{j} \frac{V_{j}}{V_{j}^{\prime}} e_{j}(\mathbf{k}) \nonumber\\
C_{3} (\mathbf{k}) = & \frac{3 H_k}{2 \sqrt{2 k^3} V'_a} e_a(\mathbf{k}) - \frac{3 H_k}{2 \sqrt{2 k^3} V'_i} e_i(\mathbf{k}) .
\end{align}
where all the quantities are evaluated at horizon crossing. 
Notice that modes with different $k$s cross the horizon at different times, so modes that cross the horizon first have more time to roll down. This effect would produce a scale dependence in the coefficient of $e_{a}$. This effect is negligible in our regime, as the fluctuations are mostly in the $e_{i}$ direction. 

In the limit of interest, $V'_a \gg V'_i$ and $V_i \gg V_a$ so that 
\begin{eqnarray}
C_1 &\simeq& -\frac{8 \pi G H_k}{\sqrt{2 k^{3}}}  \frac{V_i}{V'_i}  e_{i}(\mathbf{k}) =   - \frac{3 H_k^3}{\sqrt{2 k^3} V'_i} e_i(\mathbf{k}), \nonumber \\ 
C_3 &\simeq&  - \frac{3 H_k}{2 \sqrt{2 k^3} V'_i} e_i(\mathbf{k}),
\end{eqnarray}
which can then be plugged in Eq.~\ref{EqMaster1}.

\section{Post-Inflationary Evolution}

In the simplest scenario inflation is followed by a phase of radiation domination, where the inflaton has damped its energy into the standard model bath while the axion remains decoupled. During this phase, the axion field behaves initially as dark energy, while around $m_a \sim H$ it stars oscillating and behaves as cold dark matter. Since we are interested in parameters for which $m_a$ is typically of the same order as $H$, the axion field stars oscillating almost immediately after reheating. The axion field therefore traverses a distance much smaller than $f_a$, keeping fixed the misalignment angle. We can also safely assume that $\frac{\delta \phi_a}{\phi_a}$ remains fixed in this period as the axion is moving near the origin, in the quadratic part of the potential.

\subsection{Initial Conditions in Radiation Era}
In this section we discuss how the primordial perturbation produced during inflation become perturbations in the species that compose the cosmic fluid during the Hot Big Bang. 
After BBN, we assume that the cosmic fluid is composed of four species: photons, dark matter in the form of axions, baryons and massless neutrinos behaving approximately as fluids. 
Ignoring neutrino isocurvature modes, the quantities of interests are the fractional energy density fluctuations $\delta_j \equiv \delta \rho_j / \rho_j $ and the gravitational potential $\Phi$.
In the longitudinal gauge, for the superhorizon modes one can write the initial conditions in the form \cite{Ma:1995ey, Langlois:1999dw}:
\begin{align}
\delta_ \gamma & = - 2\Phi, \\
\delta_b = & \frac{3}{2} \delta_\nu = \frac{3}{4} \delta_\gamma, \\
\delta_a = &  S + \frac{3}{4} \delta_\gamma,
\end{align}
where we have introduced the isocurvature mode for the axion $S$, the evolution of which can be studied separately. 
These perturbations constitute the initial conditions for the computation of the effect of the isocurvature on the CMB and large scale structure. In the following we will relate these to the perturbations during inflation.

\subsubsection{Adiabatic Mode}
For the adiabatic mode, all the species have the same covariant curvature perturbation. This means 
$\frac{1}{4}\delta_\gamma = \frac{1}{4}\delta_\nu = \frac{1}{3} \delta_b = \frac{1}{3}\delta_a = \mathcal{R}$. 
In our scenario, since inflation is dominated by a single field, one gets the standard result  $\mathcal{R} \approx C_1$~\cite{Polarski:1994rz,Langlois:1999dw}, i.e.
\begin{align}
\mathcal{R} \simeq -\frac{8 \pi G H_k}{\sqrt{2 k^{3}}} \sum_{j} \frac{V_{j}}{V_{j}^{\prime}} e_{j}(\mathbf{k}) \simeq -\frac{3 H_k^3}{\sqrt{2 k^3} V'_i} e_{i} (\mathbf{k})
\end{align}
up to corrections of order $\sim V_a/V_i \ll 1$.
%In other words, the $C_3$ contribution to $\Phi$ is negligible.  

In the following we explain this result. We can compute the covariant curvature perturbation in another way, using the $\delta N$ formalism~\cite{Sasaki:1995aw, Lee:2005bb}. In the superhorizon limit, $\mathcal{R}(t)$ is equal to the difference in the number of e-foldings $\delta N(t_*, t)$ between a flat hypersurface at $t=t_*$ and a uniform density hypersurface at time $t$.
During inflation and in the slow roll regime, the number of e-foldings depends on the field trajectory as
\begin{align}
N = \int_{t_*}^t H dt \simeq \frac{1}{M_{Pl}^2}\int_{\phi_{ik}}^{\phi_{i \, end}} \frac{ V_i }{V'_i} d\phi_i.
\end{align}
Since inflation ends on a constant $\phi_i$ surface and the energy density is dominated by the inflaton field for all the relevant cosmological evolution,
inflation ends on a constant density hypersurface, so the variation of the number of e-folds involves only the field at horizon crossing\cite{Vernizzi:2006ve}
\begin{align}
\delta N = \frac{V_i}{M_{Pl}^2 V'_i} \left(\delta \phi_i\right)_k,
\end{align}
where the k subscript denotes values taken at horizon exit. We finally note that the $C_3$ component in $ \left(\delta \phi_i\right)_k$ is suppressed by ${\cal O}\left(\frac{V_a}{V_i}\right)$ (see Eq.~\ref{EqMaster1}), and therefore $\mathcal{R} \approx C_1$.

\subsubsection{Isocurvature Mode}
We can find the initial conditions for the isocurvature mode $S_a$ by taking $\Phi\to 0$. Since the contribution of $C_3$ to $\Phi$ is negligible, this is equivalent within 
 the approximation  $V_a/V_i \ll 1$ to setting $C_1 $ to zero.
 Therefore
 \begin{align}
     S = 2 \left.\frac{\delta \phi_a}{\phi_a}\right|_{C_1 \to 0} = & \frac{4}{3}\frac{V'(\phi_a)_k C_3}{\phi_{ak}} = \\ = & - 2 \frac{V'(\phi_a)_k}{\phi_{ak}} \frac{H_k}{\sqrt{2 k^3} V'(\phi_i)_k} e_i (\mathbf{k}). \nonumber
 \end{align}

\subsection{Power Spectra}
Given the results of the previous sections we can compute the various power spectra
\begin{align}\label{Eq:Spectra}
\Delta^2_{\mathcal{R}}(k) = \frac{k^3}{2\pi^2} & \left< \mathcal{R}^2 \right> = \frac{H_k^2}{8 M_{Pl}^2 \pi^2 \epsilon_k}, \\ 
\Delta^2_{S}(k) = \frac{k^3}{2\pi^2} \left< S^2 \right> & = \frac{V'(\phi_a)^2_k H_k^2}{\pi^2 V'(\phi_i)^2_k \phi_{ak}^2} = \frac{4}{9} \frac{V'(\phi_a)_k^2 }{H_k^4 \phi_{ak}^2} \, \Delta_\mathcal{R}^2, \\ 
\Delta^2_{S\mathcal{R}}(k) = \frac{k^3}{2\pi^2} \left< S \, \mathcal{R} \right> & = \frac{3 H_k^4 V'(\phi_a)_k}{2 \pi^2 V'(\phi_i)_k^2 \phi_a } = \frac{2 V'(\phi_a)_k}{3 H_k^2 \phi_{ak}} \Delta_\mathcal{R}^2,
\end{align}
where $\epsilon_k$ in the first equation is the slow roll parameter for the inflaton field. 

In the next section we will use these results to place constraints on the (low) scale of inflation from the last Planck data release. 

\section{Lower bound on inflation scale from Planck}

In analogy to Ref.~\cite{Beltran:2006sq} (see the Appendix for a comparison with this standard high-scale scenario), and using Eq.~\ref{Eq:Spectra}, we can define the isocurvature ratio
\begin{align}\label{Eq:IsoRatio}
\alpha \equiv \frac{\langle |\mathcal{S}(k)|^2\rangle}{\langle |\mathcal{S}(k)|^2\rangle + \langle |\mathcal{R}(k)|^2\rangle} \simeq \frac{4\, V_{a k}^{\prime 2}}{9 \, H_k^4 \phi_{ak}^2},
\end{align}
where we assumed $\langle |\mathcal{S}(k)|^2\rangle \ll \langle |\mathcal{R}(k)|^2\rangle$. With this ratio at hands, we can derive the bounds on our model set by the last Planck data release~\cite{Planck:2018jri}. In particular, we can use the values in Tab.~14 for the fully-correlated case which sets $\alpha_{\rm low-scale} \lesssim 10^{-3}$ for Planck TT,TE,EE+lowE+lensing. This in turn leads to a lower limit on the Hubble parameter during inflation for a given axion mass
\begin{equation}\label{Eq:BoundHubbleLow}
H \gtrsim 4.6 \, \text{eV} \, \Big(\frac{m_a}{\rm eV}\Big)\Big(\frac{10^{-3}}{\alpha}\Big)^{1/4},
\end{equation}
which also means that the scale of inflation must be larger than
\begin{equation}\label{Eq:BoundScale}
\Lambda_{\rm inf} = \Big(\frac{3 H^2 M_{\rm pl}^2}{8 \pi}\Big)^{1/4} \\ \nonumber \gtrsim 140 \, \text{TeV} 
 \,\Big(\frac{m_a}{\rm eV}\Big)^{1/2}\Big(\frac{10^{-3}}{\alpha}\Big)^{1/8},
\end{equation}
where we used $M_{\rm pl} = 1.22 \times 10^{19} \, \rm GeV$. This lower limit assumes the axion to constitute the entirety of dark matter. This condition itself, however, already sets a relation between the Hubble constant during inflation and the axion mass, as we now explain.

We can distinguish two cases, the QCD axion, for which the mass is a function of the standard model plasma temperature and evolves with time~\cite{Abbott:1982af, Preskill:1982cy, Dine:1982ah}, and the ALP case for which the mass is fixed~\cite{Hui:2016ltb}. In these two cases the relic abundance reads
\begin{eqnarray}\label{Eq:RelicAbundances}
    \Omega_a h^2  = 
\begin{cases}
        2 \times 10^{4} \left( \frac{f_a}{10^{16} \text{GeV}}\right)^{7/6}\left< \theta_{a,i}^2 \right>,  \, \text{QCD axion } \\
        0.12 \, \Big(\frac{f_a \sqrt{\left< \theta_{a,i}^2 \right>}}{1.9 \times 10^{12} \, \rm GeV}\Big)^2 \Big(\frac{m_a}{10^{-2} \, \rm  eV}\Big)^{1/2}, \, \text{ALPs}
\end{cases}
\end{eqnarray}
and we also remind the reader that the relation between the axion decay constant and its mass for the QCD axion case is $m_a \simeq 5.7 \rm \, \mu eV \, \Big(10^{12} \rm GeV / f_a\Big)$~\cite{GrillidiCortona:2015jxo}. 

In both cases, ALP and QCD axion, a small initial angle would lead to a small relic abundance. This initial angle $\theta_{\rm I}$ is the axion angle at the end of inflation, which is also a product of inflationary dynamics. In particular, during inflation, when the fields are slow-rolling, the dimensionless axion field $\theta = \phi_a/ f_a$ evolves as 
\begin{equation}
\theta(t) = \theta_i e^{\frac{-m_a^2 t}{3 H}},
\end{equation}
where $\theta_i$ is the angle at the beginning of inflation, and therefore at the end of inflation it will be 
\begin{equation}
\theta_{\rm I} = \theta_i e^{\frac{-m_a^2 N}{3 H^2}},
\end{equation}
where we defined the number of e-folds $N \equiv \rm log(a_{\rm end}/a_{\rm in})$. This is the angle which roughly defines the initial conditions for misalignment to take place afterwards. Therefore, in order not to get too small relic abundance this provides also a bound on the ratio between the axion mass and the Hubble constant during inflation. In particular, if we do not want an extremely small $\theta_{\rm I}$, say $\gtrsim 10^{-3}$, then 
\begin{equation}\label{Eq:BoundHubbleAbundance}
H \gtrsim  \, 1.2 \,  m_a \Big(\frac{N}{30}\Big)^{1/2}, 
\end{equation}
which is a slightly weaker condition than the one coming from isocurvature pertubations, Eq.~\ref{Eq:BoundHubbleLow}, but roughly of the same order. A precise comparison of the two bounds is highly model dependent. In any case, even though isocurvature bounds currently do not offer much more new information, it is exciting that future experiments may probe a scale of inflation not currently constrained by the relic abundance argument. A new discovery of a CDM isocurvature component in the primordial spectrum fully correlated with the adiabatic one, could offer a window on the dark matter conundrum. Furthermore, in the next section we show that our scenario gives a rather distinctive prediction for the spectral index of the isocurvature perturbations. 

\section{Predictions for the spectral indices}

 In the following we compute the spectral spectral indices of the two perturbations. These are defined via the power-law parametrizations
\begin{align}\label{Eq:SpectraParametrization}
\Delta^2_{\mathcal{R}}(k) &= A^2 \left( \frac{k}{k_0}\right)^{n_{ad}-1}, \nonumber \\ 
\Delta^2_{S}(k) & = B^2 \left( \frac{k}{k_0}\right)^{n_{iso}-1}, \nonumber \\ 
%\Delta^2_{S\mathcal{R}}(k) & = A B \cos\Delta_{k_0} \left( \frac{k}{k_0}\right)^{n_{cor}+\frac{1}{2}(n_{ad}+n_{iso})-1},
\end{align}
where $k_0$ is a pivot scale and the quantities $A$ and $B$ are overall normalizations determined experimentally. The spectral indices in our cases can then be derived taking a derivative respect to $k$ of Eq.~\ref{Eq:SpectraParametrization} and using the expressions in Eq.~\ref{Eq:Spectra} for the spectra in our model. The adiabatic spectral index reads
\begin{eqnarray}\label{Eq:SpectralAdiabatic}
n_{\rm ad} - 1 &=& \frac{k_0}{A^2} \left. \frac{d\Delta^2_{\mathcal{R}}(k) }{dk}\right|_{k = k_0} \nonumber \\ &=& \frac{k_0}{A^2} \frac{1}{8 M_{\rm Pl}^2 \pi^2}\left. \frac{d(H_k^2/\epsilon_k))}{dk}\right|_{k = k_0} \nonumber \\ &\simeq&    - 6 \,\epsilon_i + 2 \, \eta_i - 4 \epsilon_a \simeq 2 \, \eta_i,
\end{eqnarray}
where we took $\epsilon_i \ll 1$ and where in order to arrive to the last expression we used the following relations: $\epsilon = - \dot{H}/H^2$, $\dot{\epsilon} = H \epsilon (4 \epsilon - 2\eta)$. 

In a similar manner one gets for the isocurvature spectral index
\begin{eqnarray}\label{Eq:SpectralIso}
n_{\rm iso} - 1 &=& \frac{k_0}{B^2} \left. \frac{d\Delta^2_{\mathcal{S}}(k) }{dk}\right|_{k = k_0} \\ &=& \frac{k_0}{\pi^2 B^2}\left. \frac{d(H_k^2 V'(\phi_a)_k/V'(\phi_i)_k^2 \phi_a^2))}{dk}\right|_{k = k_0} \nonumber \\ &\simeq& 2\Big( \eta_i - \epsilon_a - \epsilon_i \Big) \sim (n_{\rm ad} - 1), \nonumber
\end{eqnarray}
where we took the axion potential to be quadratic. It is curious that while in general the two indices are not the same, they are very nearly identical for low scale inflation up to very small corrections of the order $\mathcal{O}(f_a^2/M_{\rm pl}^2)$.

\section{Conclusions}

In this \textit{Letter} we have pointed out a new source of isocurvature fluctuations for light scalar dark matter, and in particular for the axion, in the case of low-scale inflation. While, in standard set-ups, isocurvature fluctuations are due to the quantum fluctuations of the axion field during inflation, in our case the isocurvature fluctuations are a consequence of the classical rolling of the axion field along its potential. Since the duration of inflation is inhomogeneous due to inflaton perturbations, the excursion of the axion field at the end of inflation would be different in different patches, thereby creating CDM isocurvature modes. 
%When the slope of the axion potential is larger than that of the inflaton field, generically the case for low-scale inflation, then isocurvature modes are sourced. 
As a smoking gun feature, these modes are fully correlated with the adiabatic ones and their spectral indices coincide.  Their detection, together with that of axion dark matter, may then shed light on the very first moment of the universe and in particular on the physics of cosmic inflation. 

\section{Acknowledgments} 

We warmly thank Davide Racco for discussion and comments on the manuscript. AC thanks Tel Aviv University for hospitality during various stages of this work. MG is grateful to CERN-TH for hospitality during the Theory Institute ``New Physics from Galaxy Clustering", when this idea originated. MG is supported in part by Israel Science Foundation under Grant No. 1302/19. MG is also supported in part by the US-Israeli BSF grant 2018236 and NSF-BSF grant 2021779. GR is grateful for the support of the Alexander Zaks Scholarship. GR is also supported by European Research Council (ERC) under the EU Horizon 2020 Programme (ERC-CoG-2015 - Proposal n. 682676 LDMThExp). This article is based upon work from COST Action COSMIC
WISPers CA21106, supported by COST (European Cooperation in Science and Technology).

%%%%%%%%%%%%%%%%%%
%%%%%%%%%%%%%%%%%%

\bibliographystyle{bibi}
\bibliography{biblio}

\onecolumngrid
\appendix

\clearpage

\begin{center}
\textbf{\large Supplementary Material}
\end{center}
%%%%%%%%%%%%%%%%%%%%%%%%%%%%%%%%%%%%%%%%%%%%%%%%%%%%%%%%%%%%%%%%%%%%%%%%%%%%%%%
%%%%%%%%%%%%%%%%%%%%%%%%%%%%%%%%%%%%%%%%%%%%%%%%%%%%%%%%%%%%%%%%%%%%%%%%%%%%%%%

In this Supplementary Material we re-derive the axion isocurvature ratio for the standard case of high-scale inflation and provide extra details about the solutions for the perturbations equations of motion.

\bigskip

\twocolumngrid

\section{Isocurvature Bound in High Scale Inflation}

Here we review the isocurvature bound in models of inflation where inflation is driven by a scalar field with $V'(\phi) \gg V'(a)$ and $V''(\phi) \gg V''(a)$ \cite{Linde:1985yf,Lyth:1992tx,Fox:2004kb,Beltran:2006sq}. We will be interested in the regime where $m_a \ll H$, which goes from inflation until the axion field starts oscillating and behaves as cold dark matter. The axion is a light field during inflation and the field value develops superhorizon quantum fluctuations: 
\begin{align}
\left< |\delta a(k)|^2 \right> = \left(\frac{H}{2\pi} \right)^2 \frac{1}{k^3 / 2 \pi^2},
\end{align}
where $H$ is evaluated at horizon exit ($k = a H$), around $N \sim 60 $ number of e-folds before the end of inflation. 
Since the equation for the perturbations is dominated by Hubble friction for $m_a \ll H$, the perturbation of the field remains frozen until the value of $H$ drops to $m_a$ during radiation domination (notice that for the QCD axion this happens when $T\sim \Lambda_{QCD}$). 
At this point, the axion acts as a non-relativistic fluid, with $\rho_a = m_a n_a = m_a^2 \left< a^2 \right> = m_a^2 f_a^2 \left< \theta^2 
\right>$ and the power spectrum of the isocuravature mode is
\begin{align}
\langle |S(k)|^2\rangle_{\rm high-scale} = \frac{2 \, H_k^2 }{\, k^3 \, f_a^2 \langle \theta^2 \rangle},
\end{align}
The curvature power spectrum is the same as in the low-scale inflation case, therefore the isocurvature ratio becomes
\begin{align}
\alpha \simeq \frac{\Delta_S^2}{\Delta_\mathcal{R}^2} = \frac{M_{\rm Pl}^2 \epsilon_k}{\pi f_a^2 \, \langle \theta^2 \rangle}.
\end{align}
with no correlation.

\section{Solution of the Perturbation Equations} 

In the following we solve the equations for the scalar field perturbations and the gravitational potential in the superhorizon limit. The equation for the field perturbations of the inflaton (and analogously for the axion) can be written as~\cite{Polarski:1994rz}
\begin{align}
\delta \ddot{\phi} + 3 H \delta \dot{\phi} + V''(\phi) \delta \phi = 4 \dot{\phi} \dot{\Phi} - 2 V'(\phi) \Phi,
\end{align}
which in the slow roll limit reduces to 
\begin{align}
3 H \delta \dot{\phi} + V''(\phi) \delta \phi = - 2 V'(\phi) \Phi.
\end{align} 
Dividing by $3H$ 
%we have  
%\begin{align}
%\delta \dot{\phi} + \frac{V''(\phi)}{3 H} \delta \phi = %- \frac{2 V'(\phi)}{3 H} \Phi
%\end{align}
and then using the slow roll equation for the background field one gets
\begin{align}
\delta \dot{\phi} - \frac{V''(\phi)}{V'(\phi)} \dot{\phi} \, \delta \phi =  \delta \dot{\phi} - \frac{d}{dt} \left( \log\left(\frac{1}{V'(\phi)}\right) \right) \delta \phi = \\ = V'(\phi) \frac{d}{dt} \left( \frac{\delta \phi}{V'(\phi)} \right).  
\end{align}
So our equation reduces to  
\begin{align}
\frac{d}{dt} \left( \frac{\delta \phi}{V'(\phi)} \right) = - \frac{2 \Phi}{3 \, H},
\end{align}
or, integrating
\begin{align}
\delta \phi = - \frac{2}{3} V'(\phi) \left( \int dt \frac{\Phi}{H} + d_\phi \right), 
\end{align}
where $d_\phi$ is an integration constant. To fully solve the equations we now need to impose the constraint for the gravitational potential. Introducing the variable $\chi \equiv \int dt \, \frac{\Phi}{H} $ this becomes 
\begin{align}
H \dot{\chi} + 2 \dot{H} \chi = \frac{1}{M_{Pl}^2} \left( d_\phi \dot{\phi}^2 + d_a \dot{a}^2 \right),
\end{align}
and therefore
\begin{align}
\frac{1}{H} \frac{d}{dt} \left( H^2 \chi \right) = \frac{1}{M_{Pl}^2} \left( d_\phi  \dot{\phi}^2 + d_a^2 \dot{a}^2 \right). 
\end{align}
Consequently
\begin{align}
\chi = \frac{1}{ M_{Pl}^2 H^2} \int dt \, H \left( d_\phi \dot{\phi} + d_a \dot{a}^2 \right) = \\ = \frac{C_1}{2 \, H^2} - \frac{1}{3 \, H^2 M_{Pl}^2} \left( d_\phi V(\phi) + d_a V(a) \right),
\end{align} 
and finally
\begin{align}
\Phi = - C_1 \frac{\dot{H}}{H^2} - H \frac{d}{dt} \left( \frac{d_\phi V(\phi) + d_a V(a)}{V(\phi)+V(a)} \right),
\end{align}
which is indeed our master formula, Eq.~\ref{EqMaster1}, with $C_3$ = $d_a - d_\phi$.

\end{document}